\documentclass[manuscript]{aastex}

\slugcomment{To appear in Ap. J. Lett.}
\shorttitle{The most distant compact groups}
\shortauthors{Guti\'errez}

\begin{document}

\title{The most distant compact groups}

\author{C. M. Guti\'errez}
\affil{$^1$Instituto de Astrof\'\i sica de Canarias, E-38205 La Laguna, Tenerife,
SPAIN\\ $^2$Departamento de Astrof\'\i sica, Universidad de la Laguna, E-38200
Tenerife, SPAIN}
\email{cgc@iac.es}

\begin{abstract}
We present photometric and spectroscopic observations of the members of three previously cataloged
compact group (CG) candidates at redshifts $z>0.3$. These confirm spectroscopic 
redshifts compatible with being gravitationally bound structures  at redshifts 0.3112,
0.3848 and 0.3643 respectively, and then they are the most distant  CGs known with 
spectroscopic confirmation for all their members.  The morphological and spectroscopic
properties of all their galaxies  indicate early types dominated by an old population
of stars, with little star formation or nuclear activity. Most of the physical
properties derived for the three groups are quite similar to the average properties of 
CGs at lower redshifts. In particular, from the velocities and positions of the
respective members of each CG, we estimate short dynamic times. These leave open the
questions of identifying the mechanism for forming CGs continuously and the nature of the final
stages of these structures. 

\end{abstract}

\section{Introduction}

Compact groups (CGs) of galaxies are defined as relatively dense associations of four
or more galaxies  that are isolated from larger structures. Spectroscopic
observations of their members show that  typical velocity dispersions of such systems are
$\sim 200$ km s$^{-1}$. Although the presence of contaminant
interlopers  is unavoidable in some cases, the detection of hot intergalactic gas   X-ray
emission (Ponman et al.\ 1996) in 75 \% of the systems and intragroup diffuse light
(Nishiura et al.\ 2000; White et al.\ 2003; da Rocha \& Mendes de Oliveira 2005; da
Rocha et al.\ 2008) indicate that most CGs are indeed real physical associations.

Since the first catalog of CGs by Hickson (1982), many other catalogs have been built from
different surveys (Iovino et al.\ (2003); de Carvalho et al.\ (2005); Prandoni, Iovino, \&
MacGillivray (1994); Iovino (2002); Focardi \& Kelm (2002); Barton et al.\ (1996); Allam \&
Tucker 2000). Particularly relevant for this paper are the SDSS-based catalogues (Lee et al.\ (2004);
McConnachie et al.\ (2009), hereafter MC09). Those catalogs have been built using basically 
the criteria established originally by Hickson (1982)  based on the number of galaxy members
($\geq 4$) within some specific magnitude range, a given isolation criterion in order to
avoid considering CGs as transient or projected associations  of galaxies within larger
structures, and compactness. The exact values of such parameters must be selected in order to
get an optimum balance between completeness and the presence of foreground or background
contaminants. The fraction of contaminants present in a given catalog depends critically on
the value adopted for the surface brightness of the CG, as was explicitly demonstrated by
MC09. 

The SDSS-based catalogs have allowed for the first time the detection of a large number of CG
candidates at redshift $z\geq 0.2$. In particular, MC09 have compiled two different catalogs
containing 2,297 (catalog A) and 74,791 (catalog B) CG candidates, imposing  limits for 
their members of $r=18$ mag and 21 mag respectively. The mean number of galaxies per CG
candidate is 4.2 (i.e., most of the candidates have four galaxies). The fraction  of CG
candidates with a spectroscopic determination of redshift is 44 \% and 18\% for catalogs A
and B respectively. 

There are 2,062 CG candidates at estimated redshift $z>0.3$ (the most distant has 
$z= 0.46$) in MC09 (catalog B). Therefore, the MC09 catalog B provides a representative sample of candidates and allows for the
first time the exploration of the physical  conditions in CGs through the past $\sim 5$ Gyr.
For the great majority of the CG candidates, the spectra of just one
galaxy was obtained by SDSS. To reduce the degree of random projected associations, we restricted our study to the 23 systems
with at least two  members  with  concordant  determination of spectroscopic redshifts  at
$z>0.3$. Here, we present the results for three such candidates\footnote{Throughout this
paper we name the CGs as MC followed by the same number used in the MC09 catalog B.} at
redshifts 0.31 (MC4629), 0.38 (MC7697), and 0.36 (MC13069). 

This paper is organized as follows: after this introduction, Section 2 presents the basic issues of the observations and data
processing, while  the redshift estimations and the main observational 
properties are considered in Section 3. Section~~4 presents the main kinematic and dynamical properties. The main conclusions are presented in Section 4.

\section{New observations and data reduction}

Table 1 presents a summary of the photometric and spectroscopic observations presented in this
paper. Using the 2.5 m INT (ORM, La Palma, Spain), images of a field centred
in each of the three CG candidates were obtained. The images were bias subtracted, flat field
corrected and combined using IRAF\footnote{IRAF is the Image Reduction and Analysis Facility,
written and supported by the IRAF programming group at the National Optical Astronomy
Observatories (NOAO) in Tucson, Arizona.}. The images for the CG candidate MC4629 were taken
in photometric conditions, whilst during the observations of MC7697 and MC13069 there were
evidence for the presence of some high clouds. An approximate photometric calibration was done by
comparing with the corresponding SDSS images in the $r$ filter. The 1-$\sigma$ brightness
fluctuation of the resulting images is 27.0-27.5 mag arcsec$^{-2}$. 

The spectroscopic observations were taken at the 10.4 m GTC (ORM, La Palma,
Spain). Details of the telescope and the instrumental
configuration can be found on the telescope web site (www.gtc.iac.es). We used the OSIRIS
spectrograph with the R1000B grism. The observations were carried out in August 2010 in
service mode. A single position of a long slit crossing the two members of each CG candidate
with unknown redshift was taken. The spatial sampling was 0.25 arcsec/pixel. The slit width
was 1 arcsec. The nights were clear with traces of dust and light cirrus. For wavelength
calibration, we used HgAr and Ne lamps taken at the end of each night.  The stability of the
wavelength calibration during the night was checked with the main sky lines. The effective
spectral resolution was  7 \AA. The  spectra were analyzed following a standard procedure
using IRAF, which comprises bias subtraction, flat field correction, coaddition of
exposures of the same field, wavelength calibration, and extraction of the spectra. The flat
field correction was done only in the red part of each spectrum due to the low response of
the calibration lamp in the blue part (this correction is $\sim
1/200$). The spectral calibration provides a sampling of 2.07 \AA~and is quite accurate
($\sim 0.05$ \AA),  as checked through the position of the atmospheric OI line.  We used
standard spectroscopic stars from the catalog by Oke (1990) to correct for the response of
the configuration at different wavelengths. This does not provide an absolute
flux calibration owing to the conditions of the observations, the different orientation of
the slit width, etc. We did not correct for the telluric band at 7600 \AA.

\section{Membership and environment}

Figure~1 shows images of each CG candidate and the extracted spectra of
their members (for completeness, we also include  the spectra of the
galaxies observed with SDSS). The galaxies labeled 1 to 4 in each plot are those used to define and
characterize the three CG candidates. Galaxies 1 and 2 are those for which SDSS spectra exist
and were used by MC09 to estimate the redshift of the CG candidate, whilst galaxies
 3 and 4 are those targets that have been confirmed as members with the new
observations  presented in this paper. The extracted spectra have SNR $\sim$
12--18 per pixel at a spectral position around 6500 \AA~with the exception of MC4629-3
(the brightest object of the sample observed with the GTC), which has SNR$ \sim50$ per pixel.
The spectra of all the galaxies are dominated by typical galactic absorption features. A minimum number of five absorption lines was identified in the spectrum
of each galaxy; the most important being CaII H\&K, MgI, the G-band, H$_\beta$, and, in some
cases, NaI and H$_\alpha$; the presence of these  spectral features  is enough for a
reliable determination of redshift. This was done\footnote{For consistency, we determined 
the redshift also for the targets with previously
measured redshift in SDSS.} by cross correlating their
spectra with several synthetic,
 galactic, and stellar templates (Tonrey \& Davies 1979). The results from all these methods were consistent,
although the best accuracy was obtained using two stellar templates
of spectral class G and K obtained from the list provided by SDSS. The redshifts determined by this
method are the ones 
quoted in  Table~2. The $R$ factor for those determinations (Tonrey \& Davies 1979) ranges from 5 to 15. The
spectral calibration introduces an additional $\sim 3$ km s$^{-1}$ uncertainty. Table~2
lists the identification of each galaxy (col.\ 1), its equatorial coordinates
(cols.\ 2 and 3), $r$ magnitude (col. 4), $g-r$ color (col. 5), spectroscopic redshift
and statistical uncertainty (cols.\ 6 and 7), and $R$ factor (col. 8). The main conclusion
obtained from the spectra is  that all the galaxies proposed as members of the respective CG
candidates have concordant redshifts. In other words, we confirm that the three CG
candidates are real in the usual meaning of the term. Their estimated redshifts are
0.3112, 0.3848, and 0.3643 for MC4629, 7697 and 13069 respectively.

The morphological analysis of galaxies at
redshifts $\sim 0.3$ is limited from ground based images. In fact, at the distances of the
three CGs, the spatial scales are 4.53, 5.21, and 5.05 kpc arcsec$^{-1}$ for MC4629, MC7697,
and MC13069 respectively\footnote{Throughout this paper we adopt a standard lambda
cold dark matter model with $H_0=71$ km s$^{-1}$ Mpc$^{-1}$;
$\Omega_M=0.27$, and $\Omega_\Lambda=0.73$).}. Nevertheless, none of the CG candidates shows conspicuous signs of interactions and/or major distortions. 

None of the galaxies shows evidence of significant star formation or AGN activity. Only in the case
of the galaxies MC7697-3 and MC7697-5 (this object was not included as member of the CG in
MC09; however we took a spectra that allow to determine a redshift concordant with being 
a member of the CG, see below) tiny lines of $[OII](\lambda3727)$ \AA\
were detected.  So, the morphological and spectroscopic analysis of the galaxies in the
three CGs, shows that all the members are old ellipticals. 
The absence of galaxies with emission lines contrasts with what is found in Hickson
CGs. In fact, several authors (e.g., Martinez et al.\ 2010 and reference therein)
have studied samples from Hickson CGs and found that 60--70\% of all the galaxies have
emission lines. About one quarter and one third of such emission is due to AGN and
star formation activity respectively.

The criteria imposed to select the CG candidates by MC09 are basically the same as
those adopted for Hickson's sample. However, owing to the fact  that the sample of
galaxies used by MC09 is limited to those objects with $r<21$ mag, the criteria of
isolation as   defined in that paper can be only applied for those candidates as
MC4629 in which the brightest galaxy has $r<18$ mag. We have checked that extending
the sample of galaxies considered in the SDSS photometric survey up to $r=22$ mag,
neither MC7697 nor MC13069 strictly follow the isolation criteria. In the case of
MC7697 there is a galaxy about 2 mags fainter than the brightest galaxy of the
group at $\sim 22$ arcsec from the center of the group. For MC13069 there are
several nearby objects within a range of 3 mags from the brightest galaxy of the
group. We studied also the density of objects in the near environment of each of
the three groups. Fig. 2 presents the projected density of galaxies in $r$-band
SDSS images (the results obtained are similar analyzing the INT images) as a
function of the distance to each CG candidate. Only objects catalogued as galaxies
in SDSS with magnitudes $r<22$ and photometric error $<0.2$ mag have been
considered for that plot.  The mean
densities of galaxies at distances
between 1 and 2 Mpc from each candidate, are 43, 38 and 46 gal Mpc$^{-2}$ for MC
4629, MC7697 and MC13069 respectively. The members of MC4629 and MC13069 seem
to be immersed in a overdense region of $\sim$1 Mpc. So, two out of the
three CGs analyzed here, are embedded in a rich environment. These results, although
not statistically significant, are compatible with the findings by Mendel et al.
(2011) in their analysis of the MC09 (catalog A).

A brief description of the specific properties of each group is presented below:

\begin{itemize}

\item MC4629: Galaxies MC4629-1, MC4629-2, and MC4629-3 have absolute magnitudes in
$r$ in the range $-22$ to $-23$ mag.  The projected separation between MC4629-1 and
MC4629-3 is $\sim 5$ arcsec ($\sim 23$ kpc), whilst their difference in radial
velocity is 420 km s$^{-1}$. Although  there are no significant morphological
distortions, both galaxies are slightly elongated and seem to be immersed in a
halo   with an extension of $\sim 90$ kpc at least that comprises also the galaxy
MC4629-4. There is an extended X-ray source  (Vikhlinin et al.\ 1998) at 20 arcsec
($\sim 90$ kpc) from the optical center of the CG, and having a core radius of
$34\pm 13$ arcsec and a flux  $(12.1\pm 2.8)\times 10^{-14}$ erg s$^{-1}$ cm$^{-2}$
that corresponds to  a luminosity of $3.5\times 10^{43}$ erg s$^{-1}$. That X-ray
luminosity is higher than the values measured in local CGs and  is in the typical
range of clusters of galaxies. 

\item MC7697: There is more diffuse material in this group than in MC4629. Galaxies MC
7697-1 and MC7697-3  are separated by $\sim$6 arcsec ($\sim$30 kpc). The image shows some
evidence of distortion of the outer isophotes of both galaxies and the  existence of a bridge
of light connecting them and extending out towards NW. We interpret this as
indicative of interaction between those two galaxies. For the object labeled 5 (see Fig.\
1$b$)
we determined a redshift of 0.38525, thus confirming that the object is also a member of the
group. Objects MC7697-2 and MC7697-5 seem to form part of a group of four objects inmersed
in a region  $\sim 60$ kpc $\times 45$ kpc. There is a cataloged SDSS cluster (Goto et al.\
2002) at 0.151 arcmin for which these authors identified 11 objects as potential members.

\item MC13069: The group is dominated by galaxies MC13069-1 and MC13069-2 which are
separated by  $\sim 20$ kpc; however, this proximity does not produce major distortions apart
from an asymmetry in galaxy MC13069-1 through the NE. The velocity dispersion
(Table~3; col. 3) of this group
is within the range found for massive clusters of  galaxies. This high velocity dispersion is
basically due to galaxy MC13069-2. The absence of major distortions might  indicate
that this galaxy is experiencing the
first stages of interaction with MC13069-1.

\end{itemize}

\section{Physical properties}

The radial velocity dispersion, their random errors, and the intrinsic three
- dimensional velocity dispersion (Hickson et al. 1992) are listed in Table~~3; cols
\ 3 - 5. To estimate masses of large structures, several estimators have been proposed
(Heisler at al.\ 1985) and can be applied to the specific case of CGs (Hickson et
al.\ 1992). Here, we only compute the virial masses (Table~3; col.\ 8); of course that
estimate is only reliable when the group is in virial equilibrium. As expected from the
velocity dispersions, the masses
found for the CGs  MC4629 and MC7697 are within the range of virial masses of
local  CGs estimated by Hickson et al.\ (1992); whilst MC13069 has a similar mass
to that found in rich clusters of galaxies.

 The velocity dispersion and masses obtained for MC13069 are consequence of the difference
 in velocity between this galaxy and the other members of the group; if we exclude  this galaxy from such
estimations\footnote{The remaining three galaxies would not obviously fit  the criteria for
 a CG.}, the values obtained are  281 km s$^{-1}$ and $\log (M)$ = 46.03 for the
velocity dispersion and mass respectively. Both values would fall  within the
expected range of values for CGs.

The crossing times (Table 3; col. 7) are quite short (a few hundredths of the age
of the Universe), and below $\sim 1/20$ of their cosmological distance. 
The dynamic crossing times $\log H_ot$ for the Hickson sample 
are within the range $-2.96, +0.94$, with the majority of the CGs having values within
$-2.5$ and $-1.5$. The median of that sample was  $\log H_ot=-1.80$\footnote{Those authors
used $H_o=100$ km s$^{-1}$Mpc$^{-1}$; correcting to the value of the Hubble constant
adopted in this paper, we get $\log H_ot=-1.95$}. The $\log H_ot$ values estimated for
our three groups are $-1.76$, $-1.86$, and $-2.32$, which are within the range spanned by most
of the CGs in the Hickson sample. The values found for the three CG analyzed in this paper 
 agree also with those found by Pompei et al.\ (2006) in their study of several CGs at
redshift $\sim$0.12.

These short crossing times seem to indicate that, unless a mechanism for stabilization
and/or acquisition of new members is established, the three CGs presented here could not have lasted
until the present, and so the population of CGs at redshifts $\sim$0.3 is different from the
local one. This raises the issue of the generation and fate of CGs. The most natural
scenario (Mendes de Oliveira 2006) would be the transformation of CGs into a relatively  isolated single
giant  elliptical galaxy  leaving as an additional tracer the presence of an extended
gas halo; these properties match those found for the so-called elliptical X-ray
overluminous galaxies or fossil groups. In such systems there is an absence of
intermediate bright galaxies; if fossil groups are the final stage of CGs, the lack of
intermediate bright galaxies might reflect not only the signature of an evolved stage
but also the isolation criteria adopted to define CGs. It is worth  mentioning also that the three CG candidates match the trend 
between masses and crossing times found in Hickson's sample in the sense of faster
evolution for more massive CGs.

Hickson et al.\ (1992) found weak evidence of anticorrelation between crossing times and the
difference in brightness between the two brightest galaxies of each group. Such correlations
seem sensible in a hierarchical framework.  These  differences in magnitudes between the two
brightest ranked galaxies in  our groups (in the $g$ band and after $K$ corrections) are
+0.32, +0.07,  and +0.07 for MC4629, MC7697, and MC13069 respectively. These values are
smaller than the ones found in Hickson CGs (see Fig.\ 6 in Hickson et al.\ 1992); whether this
is a consequence of selection bias or reflects some physical characteristic of the CGs will
require further investigation. The range in $g-r$ color spanned for the members of each group
are 0.85, 0.51 and 0.17 mag for MC4629, MC7697 and MC13069 respectively. These seem to
indicate a more evolved evolutive stage of MC13069. 

The range in density, $\log \rho$, spanned for the majority of CGs in the Hickson sample
ranges from -25.7 to -22.7, with a mean value of -24.16  g cm$^{-3}$, so the values found for
the three groups analyzed in this paper are within that range. The mass-to-light ratio found
for  MC13069 is typical of the values found in local CGs; however, for MC4629 and MC7697
the estimated mass-to-light ratios are a bit low as compared to most of the values found in
Hickson's sample and are more typical of those found in single galaxies, which seems to
indicate that there is not much intergalactic mass in those two groups. 

\section{Conclusions}

\begin{itemize}

\item New observations complemented with available SDSS spectra for all their galaxy members,
confirm three candidates as real CGs at redshifts $z=$ 0.3112, 0.3848,
and 0.3643 for MC4629, 7697, and 13069 respectively. Then, they are the three most
distant CGs known with spectroscopic confirmation of all their members. 

\item The estimates for the velocity dispersion of MC4629 and 7697 are 
similar to values found  for local CGs, whilst MC13069 has a dispersion 
typical of a relatively massive cluster of galaxies. This is mainly a consequence of one of the
galaxies of this group, which has a difference in velocity with respect to the
velocity of the remaining three galaxies of $\sim 1,500$ km s$^{-1}$. 

\item The galaxy members of the three CGs show an early type morphology. This is
confirmed by the spectra, which seem to be dominated by an old stellar population. Two of the CGs
(MC4629 and 13069) are immersed in more extended overdense regions.

\item The main physical properties derived from the photometry and spectroscopy are
very similar to those found in CGs at lower redshifts. The evolution as
measured from the dynamic times is relatively fast in the three cases and thus
supports a scenario with continuous formation of CGs. 

\item Statistical conclusions will require the detection and
characterization  of a larger sample of CGs; nevertheless the observation presented here points
to a scenario in which CGs are relatively transient structures that possibly collapse
into single large elliptical galaxy in an environment similar to those found in fossil groups.

\end{itemize} 

\acknowledgments 
We are particularly grateful to E. D\'\i az-Gim\'enez, who read this manuscript and gave many
useful comments and suggestions. We thank to R. Barrena who made the observations at the INT.
This paper is based on observations made with the INT and GTC telescopes (ORM, La Palma,
Spain). We have
used the  databases SDSS (http://www.sdss.org/) and
NED (http://nedwww.ipac.caltech.edu/).

\clearpage

\begin{figure}
\includegraphics[angle=0,scale=0.46,]{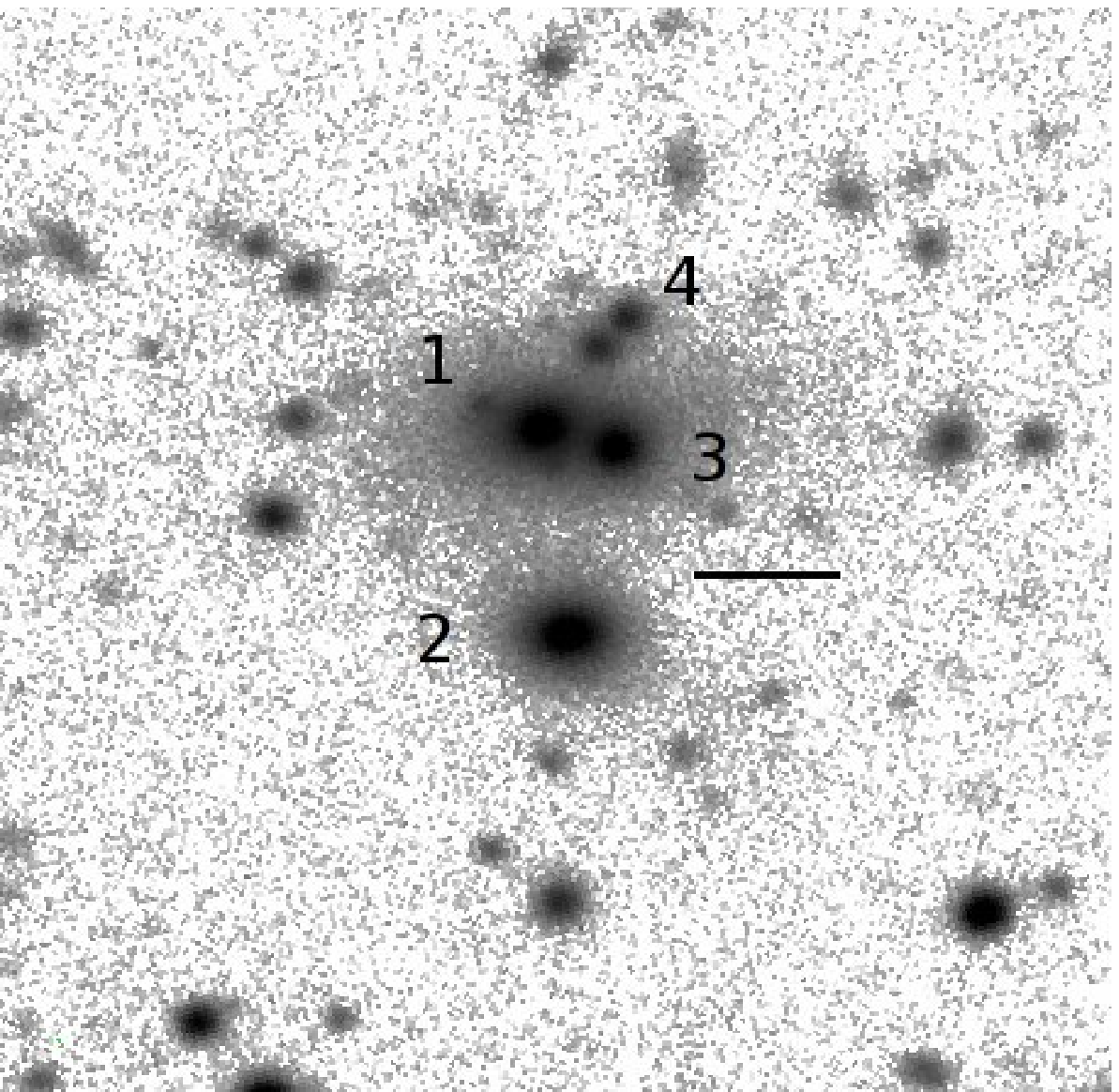}
\includegraphics[angle=0,scale=0.275]{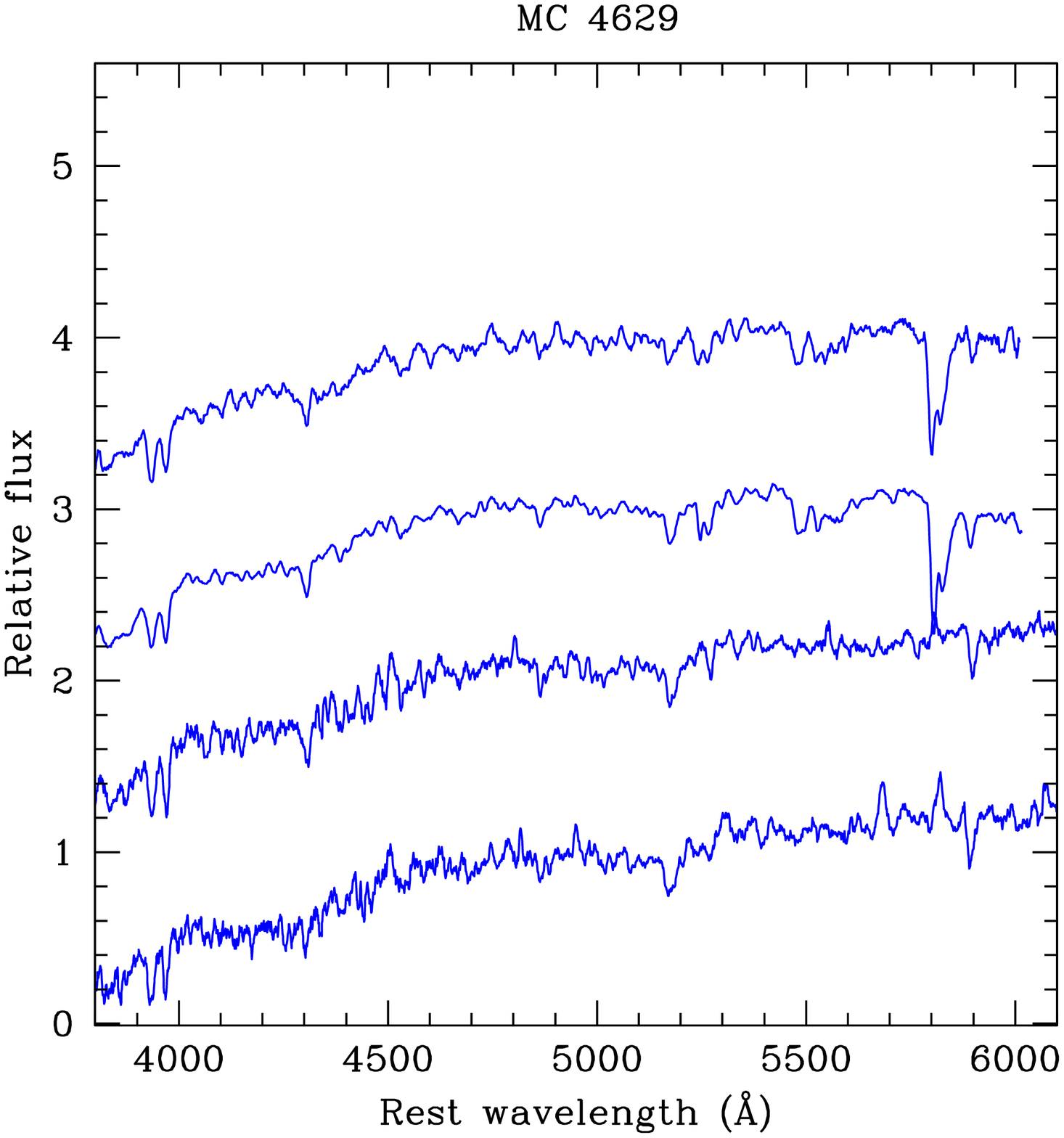}
\\
\includegraphics[angle=0,scale=0.46]{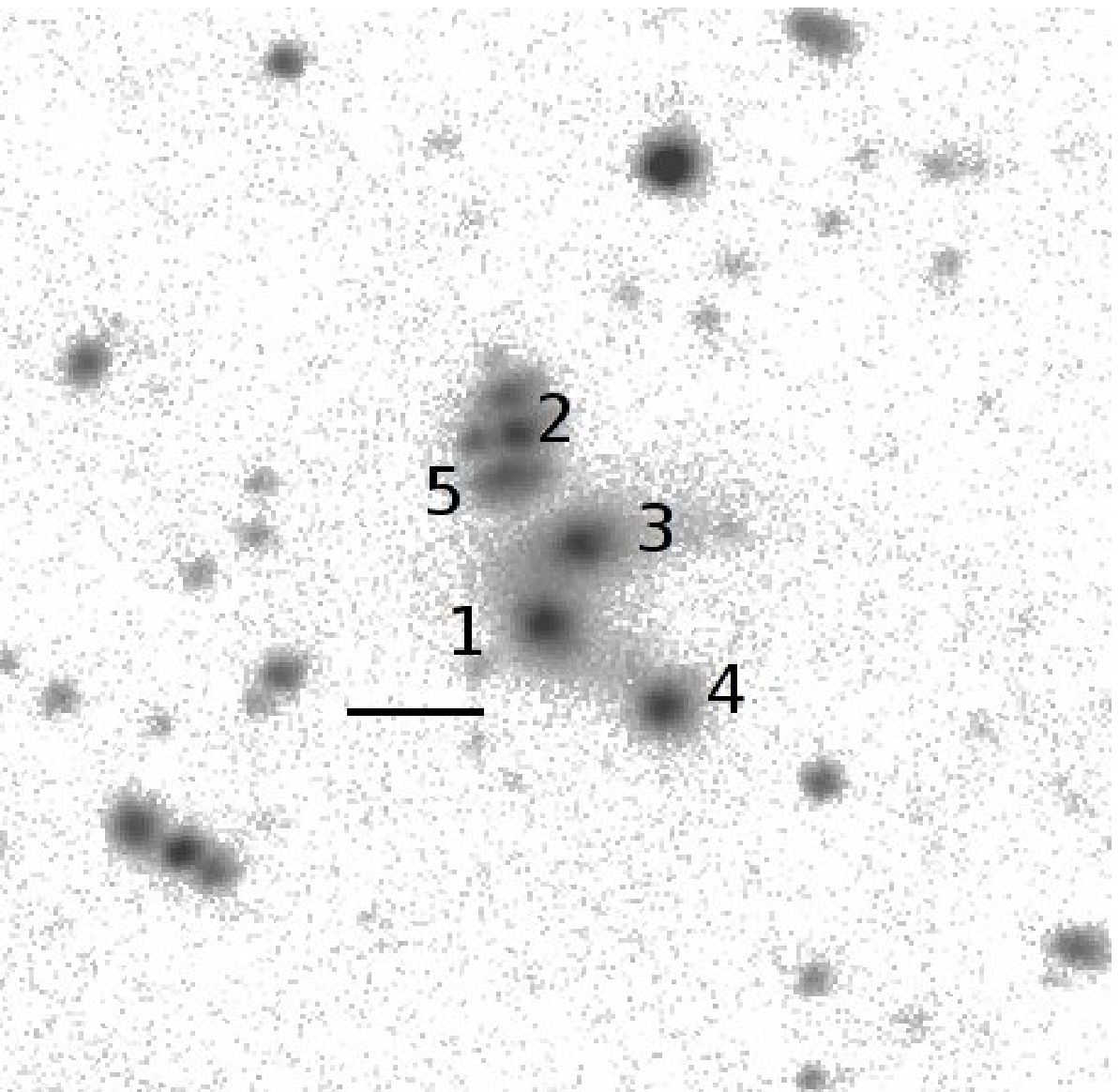}
\includegraphics[angle=0,scale=.275]{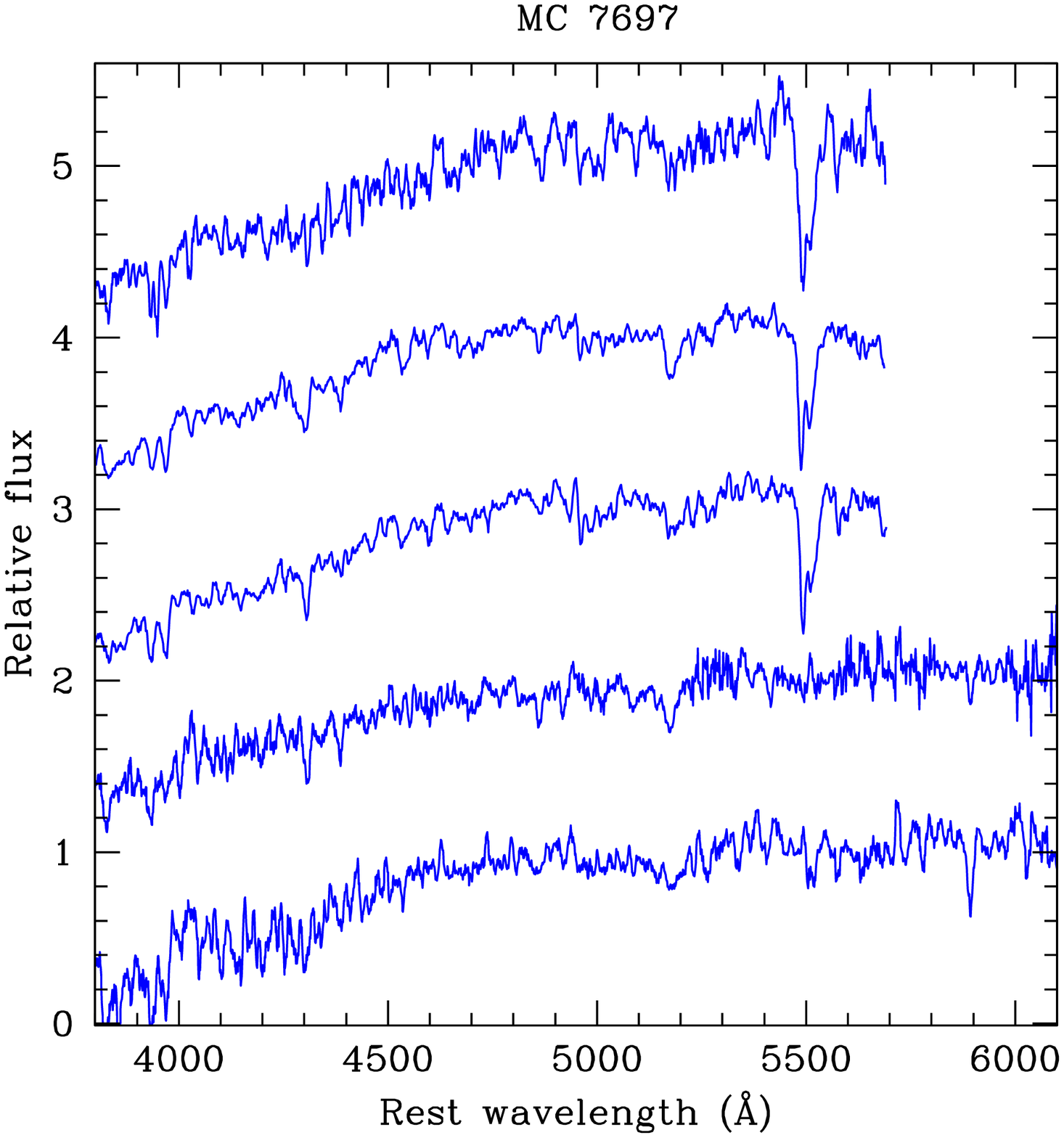}
\\
\includegraphics[angle=0,scale=0.46]{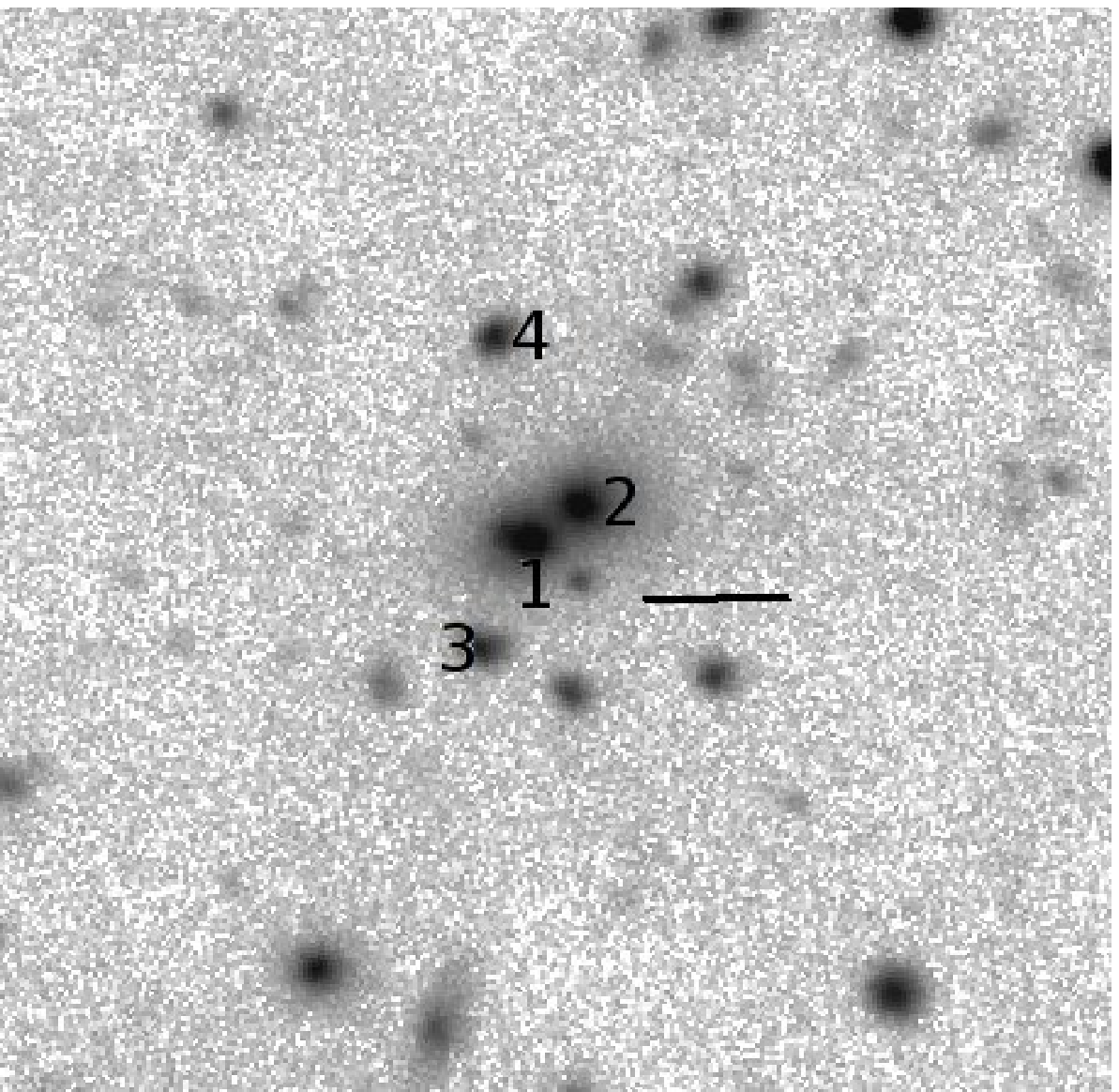}
\includegraphics[angle=0,scale=.275]{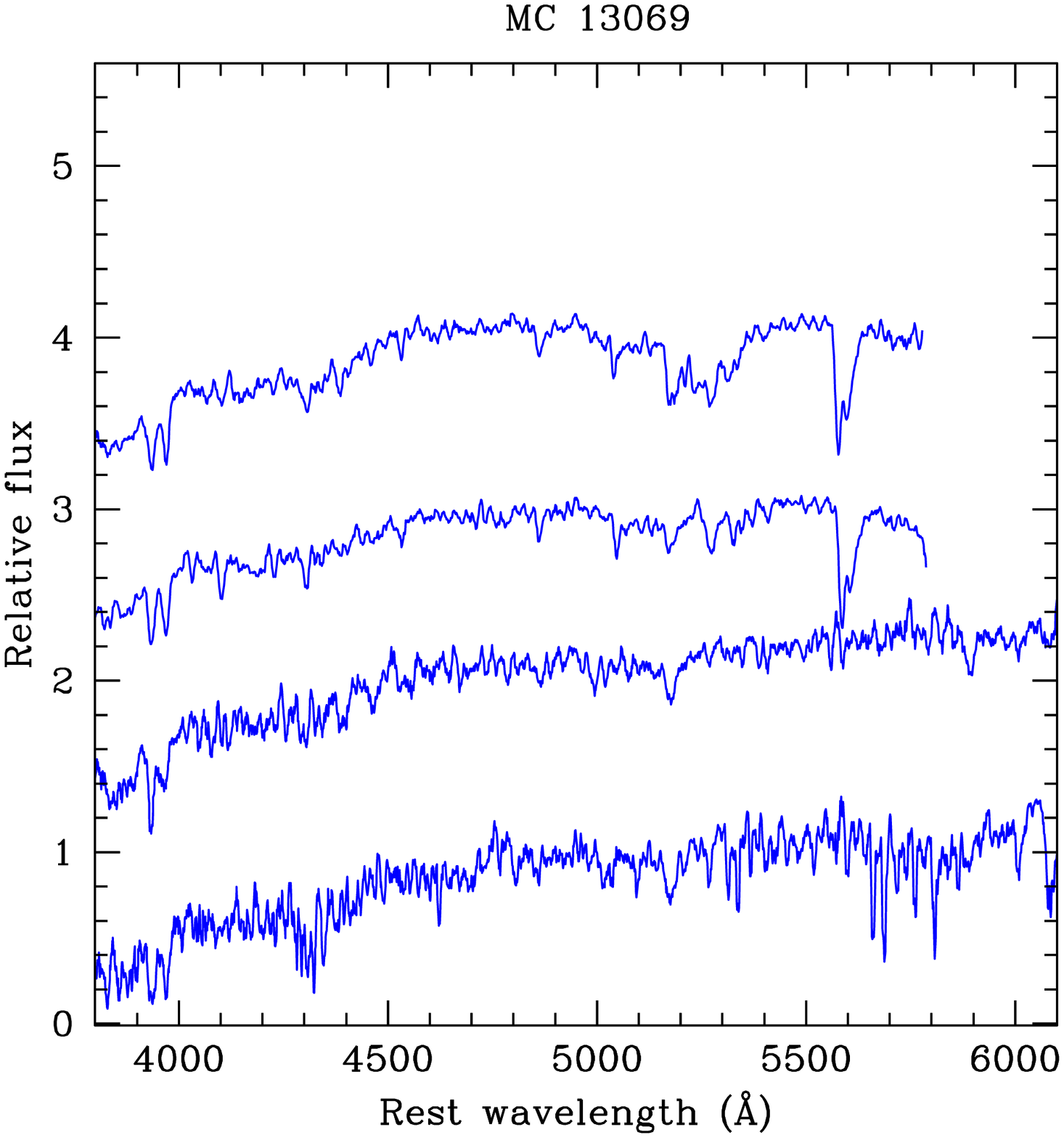}
\caption{{\it (Left:)} Image of the compact group candidates (from top to
bottom)   MC4629, MC7697, and MC13069. The horizontal line in the images
corresponds to 10 arcsec. {\it (Right:)} Rest-frame spectra of their galaxy
members. The spectra have been normalized with respect to the value at 5000
\AA. In each panel, the spectra are plotted starting from galaxy \# 1 at the
bottom. For visual purposes, all the spectra (apart from the one of galaxy \# 1) have been shifted one unit 
with respect to the precedent galaxy. The spectra of galaxies \#1 and \#2 were
obtained by SDSS.}
\end{figure} 

\clearpage
\begin{figure}
\includegraphics[angle=0,scale=0.85]{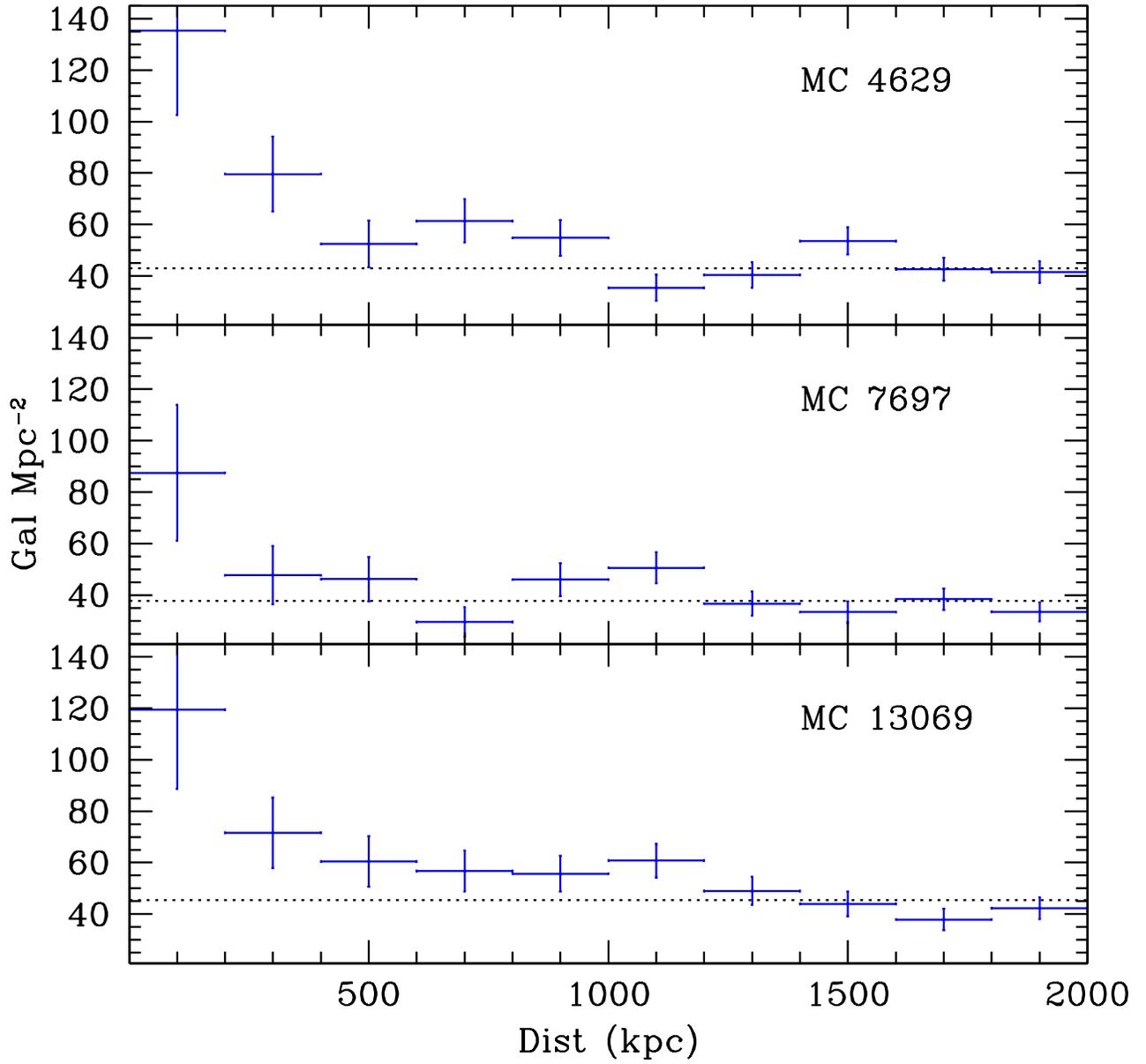}
\caption{Projected density of galaxies in a region centered on each CG candidate.
The dotted lines indicate the mean density of galaxies at distances between 1 and 2
Mpc from each group.}
\end{figure}

\begin{table}
\begin{center}
\caption{Observing log}
\begin{tabular}{lcccccccc}
\tableline
\tableline
 &\multicolumn{4}{c|}{Spectra} & \multicolumn{4}{c}{Images} \\
ID & Date & $t_{exp}$  & Seeing & Airmass & Date & $t_{exp}$ & Seeing & Airmass\\
 & (ddmmyy) & (s) & ('') & &(ddmmyy) & (s) & ('')\\
\tableline
MC4629  & 040810 &3x1200 & 0.9 & 1.51-1.87  & 240410 &3x1200 & 1.2-1.5 &
1.12-1.28 \\
MC7697 &  090810 & 3x200 & 1.0 & 1.61-1.67 & 250410 &3x1200 & 1.0-1.2 &
1.35-1.64\\
MC13069 &  150810 & 3x700 & 0.8 & 1.22-1.29 & 250410& 3x1200 & 1.0-1.2 &
1.16-1.29 \\
\end{tabular}
\end{center}
\end{table}

\begin{table}
\begin{center}
\caption{Members of each compact group}
\begin{tabular}{cccccccc}
\tableline
\tableline
ID & RA (J2000) & Dec. (J2000)  & $r$ & $g-r$ & z & $\delta$v & R\\
& & & & & & (km s$^{-1}$) \\
\tableline
MC4629-1&16    06 43.8&  23   29 15.7&  17.86&  1.80&  0.31172 & 51 & 8\\ 
MC4629-2&16    06 43.9&  23   29 01.7&  18.28&  1.58&  0.31141 & 60 & 6\\  
MC4629-3&16    06 43.5&  23   29 13.4&  18.91&  1.32&  0.31032 & 43 &12\\
MC4629-4&16    06 43.3&  23   29 21.8&  20.34&  2.17&  0.31142 & 41 &10\\   
\tableline
MC7697-1 &  14    17 53.4  &  01   14 48.9 &  18.90  &  1.62   &  0.38419 & 55& 6\\ 
MC7697-2 &  14    17 53.5  &  01   15 01.8 &  19.10  &  1.50   &  0.38417 & 70& 6\\
MC7697-3 &  14    17 53.2  &  01   14 54.0 &  19.12  &  1.11   &  0.38478 & 33&12\\
MC7697-4 &  14    17 52.9  &  01   14 42.7 &  19.53  &  1.42   &  0.38599 & 30&15\\
\tableline
MC13069-1 & 15    39 10.9  &  45   33 11.9 &  18.65  &  1.73   &  0.36352 & 53& 5\\
MC13069-2 & 15    39 10.5  &  45   33 14.0 &  18.72  &  1.74   &  0.36814 & 58& 6 \\
MC13069-3 & 15    39 11.3  &  45   33 04.9 &  20.50  &  1.64   &  0.36176 & 32&13\\
MC13069-4 & 15    39 11.0  &  45   33 25.9 &  20.63  &  1.57   &  0.36392 & 35&12\\
\end{tabular}
\end{center}
\end{table}

\begin{table}
\small
\begin{center}
\caption{Dynamic properties of each compact group}
\begin{tabular}{cccccccccc}
\tableline
\tableline
ID & z & $\sigma$ & $\delta \sigma$ &V & R & $\log$ (H$_0$t) & $\log$ (M)  & $\log$ (M/L) & $\log$ $(\rho)$ \\
   &   & (km s$^{-1}$) & (km s$^{-1}$) & (km s$^{-1}$) &(kpc) & & (g) &  (solar units) &  (g cm$^{-3})$\\
\tableline
MC4629 &0.3112 & 160 & 25 & 263 & 50 & -1.76 & 45.45 & 0.50 & -24.74 \\	    
MC7697 &0.3848 & 222 & 25 & 374 & 57 & -1.86 & 46.07 & 1.13 & -24.29 \\       
MC13069&0.3643 & 703 & 23 & 1214& 64 & -2.32 & 47.02 & 2.16 & -23.49 \\	     
\end{tabular}
\end{center}
\end{table}

\end{document}